\documentclass{article}[12pt]
\usepackage{epsf}

\usepackage{graphicx}
\title{
\vspace{-1cm}
\begin{flushright}
{\normalsize \bf UK/05-06} 
\end{flushright}
\vspace{1cm}
Charmonium Spectrum from Quenched QCD with Overlap Fermions}

\author{
S. Tamhankar${}^{a,b}$\footnote{e-mail: stamhankar01@gw.hamline.edu}, A. Alexandru${}^{a}$, Y. Chen${}^{a,c}$, S.J. Dong${}^{a}$, T.
Draper${}^a$, \\
I.  Horv\'{a}th${}^a$, F.X. Lee${}^{d,e}$, K.F. Liu${}^{a}$,
N. Mathur${}^{a,e}$ and J.B. Zhang${}^f$  \vspace{0.5cm}\\
$\chi$QCD collaboration \vspace{0.5cm}\\
${}^a$Department of Physics and Astronomy, 
University of Kentucky, Lexington, KY 40506, USA\\
${}^b$Department of Physics, Hamline University,
St. Paul, MN 55104, USA\\
${}^c$Institute of High Energy Physics,
Chinese Academy of Sciences, Beijing 100039, P. R. China\\
${}^d$George Washington University, Washington, DC 20052 USA\\
${}^e$Jefferson Lab, 12000 Jefferson Avenue, Newport News, VA 23606,
USA\\
${}^f$CSSM and Department of Physics,
University of Adelaide, Adelaide, SA 5005, Australia\\
}
\date{}

\begin{document}

\maketitle

\begin{abstract}
We present the first study of the charmonium spectrum using overlap
fermions, on quenched configurations. 
Simulations are
performed on $16^3 \times 72$ lattices, with Wilson gauge action at
$\beta=6.3345$. 
We demonstrate that we have discretization errors under control
at about 5\%. 
We obtain 88(4) MeV for hyperfine splitting using $r_0$
scale, and 121(6) MeV using the ($1\bar{P}-1\bar{S}$) scale.
This paper raises the possibility that the discrepancy between the
lattice results and the experimental value for charmonium hyperfine
splitting can be resolved using overlap fermions to simulate the charm quark
on lattice.
\end{abstract}

\section{Introduction}

Over the last few years, numerical simulations of chiral fermions have
matured. The stage of testing has passed for simulating valence chiral
fermions, and physically relevant results have been reported in
lattice simulations. All the studies so far have concentrated
on simulating light quarks. This is natural, as chiral symmetry plays an
important role for small quark masses. However, the use of overlap fermions
to simulate heavy as well as light quarks has been suggested
in~\cite{liu02}. In this paper we want to make the
point that overlap fermions can also alleviate some problems related to
simulating heavy quarks. Here we present the first quantitative study of
a heavy quark system using overlap fermions. This opens the door for the
simulation of experimentally more interesting heavy--light systems.
Using the unequal mass Gell-Mann-Oakes-Renner relation as the
renormalization condition, the renormalization factor in the heavy-light
current can be determined non-perturbatively to a high precision for
overlap fermions.~\cite{liu02} This is important for computation of heavy-light decay
constants. 

We demonstrate the value of overlap fermions to simulate heavy quarks
using hyperfine splitting in the charmonium system.
It is known that with staggered quarks, there is an ambiguity about
Nambu-Goldstone (NG) and non-NG modes for the $\eta_c$,
resulting
in widely different estimates of hyperfine splitting --
51(6)~MeV (non-NG) and 404(4)~MeV (NG)~\cite{aoki}. NRQCD converges only
slowly for charm~\cite{NR}. Including
$\mathcal{O}(v^6)$ terms
changed the result from 96(2)~MeV to 55(5)~MeV. Wilson fermions have
$\mathcal{O}(a)$
errors. Hyperfine splitting is very
sensitive to the coefficient of the correction term, $c_{SW}$.
There are many studies \cite{allton,okamoto,choe} using Wilson type valence quarks,
including some with
non-perturbative $c_{SW}$, and with continuum extrapolation. The quenched clover estimate of hyperfine
splitting has stabilized
around 70--75 MeV using $r_0$ scale~\cite{okamoto,choe,chen}, and a higher number
of about 85 MeV using ($1\bar{P}-1\bar{S}$) scale~\cite{okamoto} . Results from a 2+1 dynamical simulation using tree-level
$c_{SW}$ still
fall short of the experimental value by about 20\%~\cite{dyn}.

Although costly to simulate, overlap fermions~\cite{neuberger} have
the following desirable features:
\begin{itemize}
\item{Exact chiral symmetry on the lattice.}
\item{No additive quark mass renormalization.}
\item{No flavor symmetry breaking.}
\item{No $\mathcal{O}(a)$ error.}
\item{The $\mathcal{O}(m^2a^2)$ and $\mathcal{O}(\Lambda_{\rm QCD}ma^2)$
errors are also small, from dispersion relation
and renormalization constants.}
\end{itemize}

\begin{figure}
\begin{center}
\includegraphics[width=10cm]{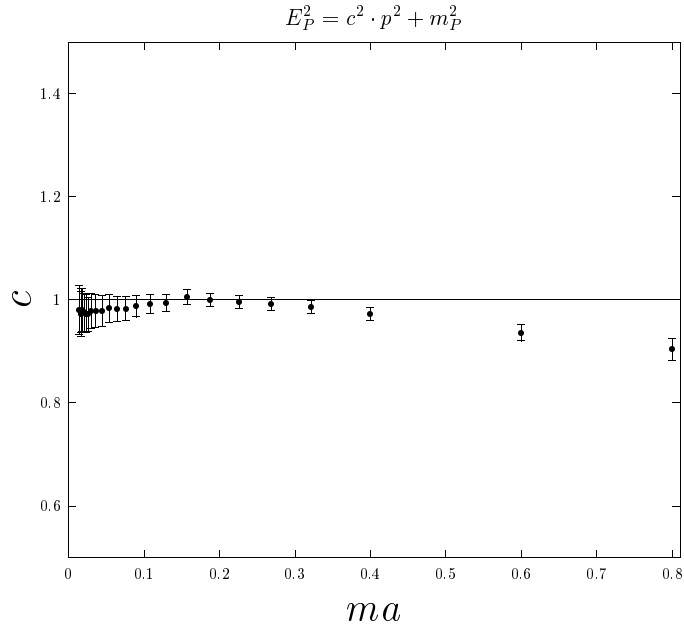}
\end{center}
\vspace{-1.0cm}
\caption{This is a plot of the speed of light, $c$,
obtained from the dispersion relation. It can be seen that the
discretization errors are only a few percent till $ma \approx
0.5$. This data comes from a $16^3 \times 28$ lattice at a spacing of
0.20fm.
 \label{fig-disp-shaojing}}
\end{figure}

\begin{figure}
\begin{center}
\includegraphics[width=12cm]{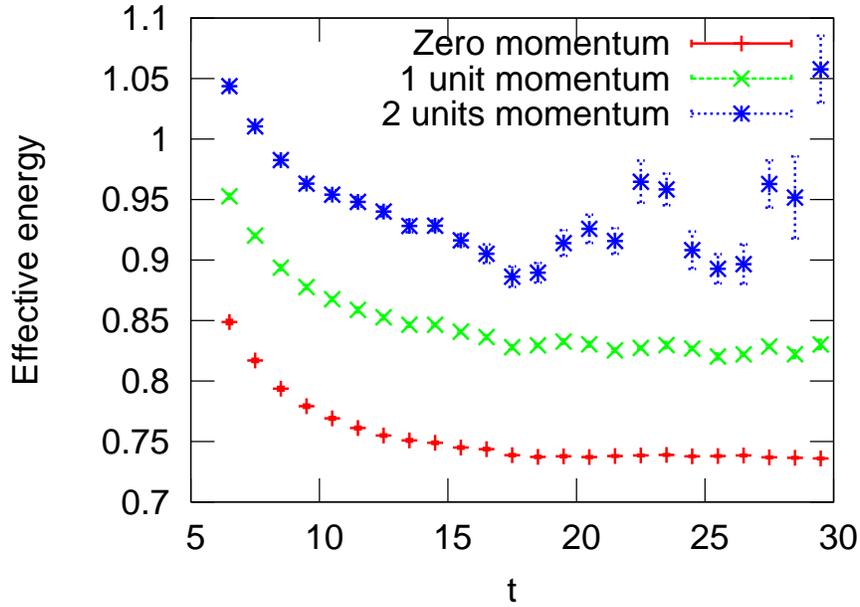}
\end{center}
\caption{Effective energies for pseudoscalar mesons, for 0, 1 and 2 units of
lattice momentum, from the $16^3 \times 72$ lattice, at $ma = 0.350$. The effective energy for 2 units of momentum
is very noisy, as explained in text. \label{fig-eff-disp}}
\end{figure}

\begin{figure}
\begin{center}
\includegraphics[width=12cm]{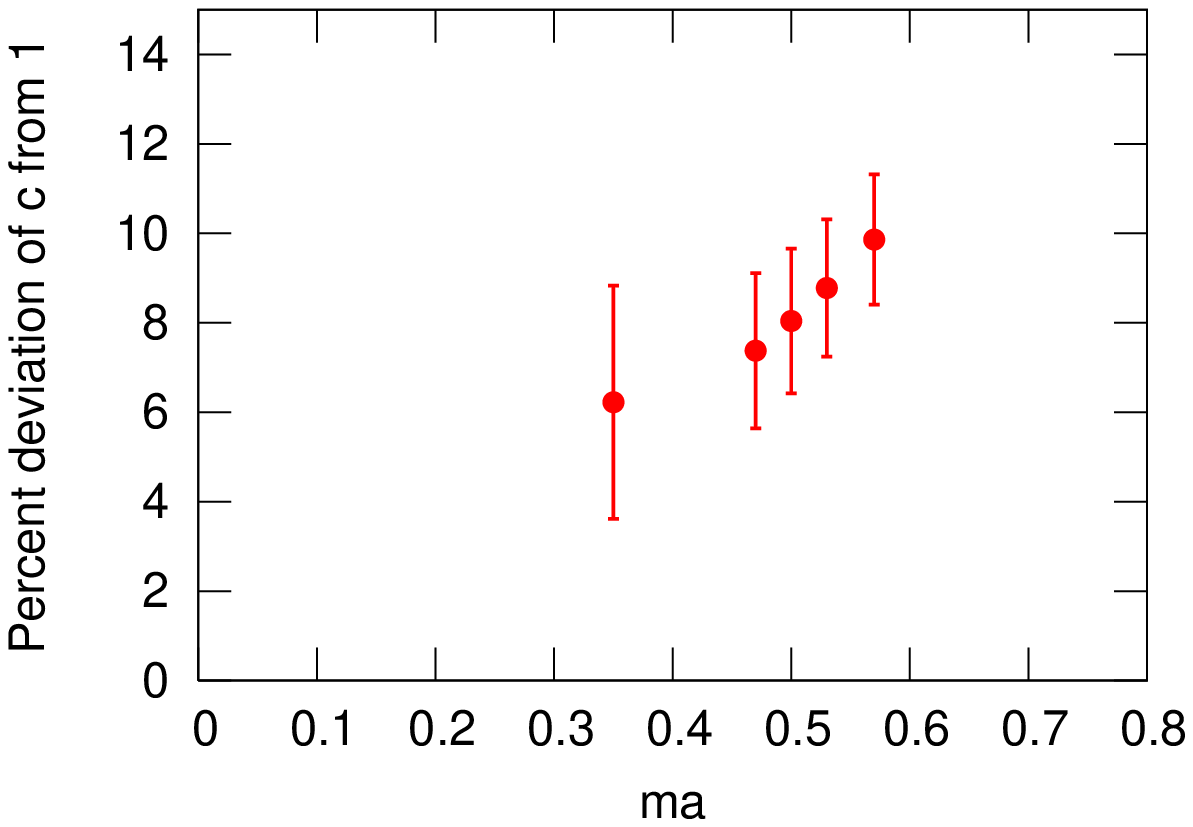}
\end{center}
\caption{Percent deviation of the speed of light from unity, as a
function of $ma$. This serves as an estimate of the percent
discretization error. Near our charm mass, the discretization errors are
about 5\%. \label{fig-disp-charm}}
\end{figure}

The first two features are especially significant for light quarks.
Many exciting results at low quark masses have been reported using
overlap fermions~\cite{overlap}. The last three features are more important for
computing charmonium hyperfine splitting using overlap fermions.
The last
feature, demonstrated in~\cite{liu02},
is an unexpected bonus in this regard. The key observation is that the
discretization errors are only about 5\% all the way up to $ma \approx
0.5$. In Fig.~\ref{fig-disp-shaojing} we reproduce a plot of the speed
of light, obtained from the pseudoscalar meson dispersion relation, as a function of $ma$ from Ref.~\cite{liu02}. This is obtained
using a $16^3 \times 28$ lattice at a spacing of
0.20~fm. 
It is harder to study the dispersion relation
on the configurations we use for this paper, because on the small volume lattice
box we use, one unit of momentum corresponds to 
about 1.6 GeV. This is a
huge amount of momentum, and as a result, the data is noisier. The
effective energies for 0,1 and 2 units of lattice momentum are shown in
Fig.~\ref{fig-eff-disp}. There is no clean plateau already for 2
units of momentum. This results in a large error bar for the energy
corresponding to that momentum. Fig.~\ref{fig-eff-disp} corresponds to
$ma=0.35$. For smaller values of $ma$, the data is even more noisy, and
it is hard to obtain the speed of light reliably for smaller masses.
However, it is expected that the deviation of speed of light from 1 is
larger for higher values of $ma$. 
Fig.~\ref{fig-disp-charm} shows percent deviation of the speed of light
from unity,
obtained from a fit to the dispersion relation as a function of quark
mass using the equation 
\begin{equation}
(E(p)a)^2 = c^2 (pa)^2 + (E(0)a)^2.
\end{equation}
It is clear from this figure that we
have discretization errors under control at about the 5--7\% level near the charm
mass, which is near $ma \approx 0.35$.

\section{Simulation Details}

Our simulations are performed on $16^3 \times 72$ isotropic lattices. We
present results on 100 configurations. The Wilson gauge action is used
at
$\beta$ = 6.3345. 
We use a multi-mass inverter to obtain propagators for  26 masses 
ranging from 0.020--0.85 in lattice
units.  Only five of these masses in the range 0.25--0.50 are used for this study.

\begin{figure}
\begin{center}
\includegraphics[width=12cm]{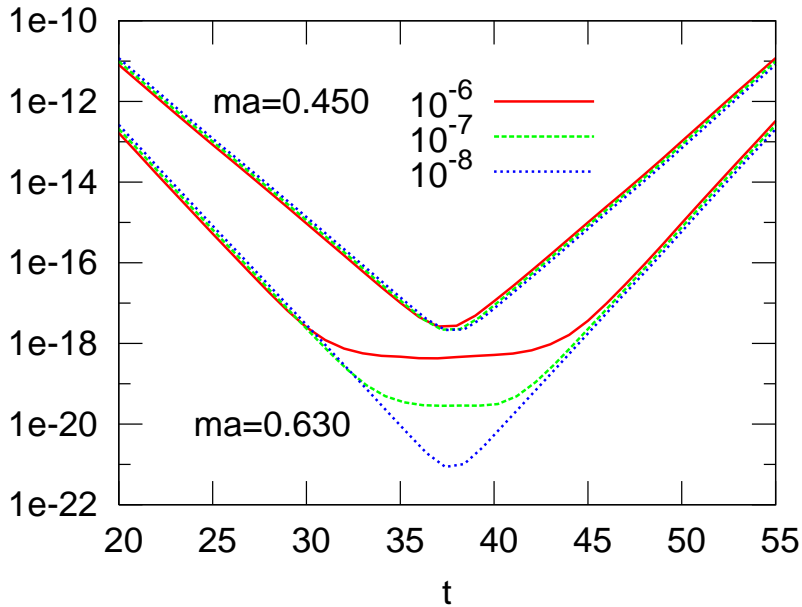}
\end{center}
\caption{Effect of inner loop precision on pseudoscalar propagators for
heavy
quarks. We study output of one spin and one color for a single
configuration for this illustration. Curves are slightly shifted
horizontally for
clarity.  \label{fig-inner}}
\end{figure}

Since overlap simulations are computationally expensive, it is important
to choose the required residues carefully -- blindly requiring
extremely precise inversions is not the optimal use of computing resources.
For overlap simulations, there are three relevant numbers: residue for
eigenvectors projected out to reduce the condition number of the matrix
to be inverted in the inner loop, residue for inner loop which computes the overlap
operator, and residue for the outer loop which actually computes the
quark propagators. For the lattices we use, we only need to project out
about 15 eigenvectors, so we simply demand a very small residue,
$10^{-10}$ for this step. Unlike this step, however, the inner and outer
loop residues demanded affect the computational cost substantially. To
determine what residue is good enough, we repeat quark propagator
inversion for one spin, one color and one configuration, and compare the
``pseudoscalar'' two-point function for various quark masses. This is not a
physical quantity since no trace over spin and color is performed, and
no configuration average is taken -- we are simply studying precision
issues here. Comparing results for inner loop residue of $10^{-6}$ with
those from inner loop residue of $10^{-7}$, we find no change for small
quark masses.  However for heavy quarks, the two-point function
falls through many orders of magnitude, and becomes very small at the
center of the lattice. To get this precisely, we find we need a small inner
loop residue -- $10^{-6}$ is not sufficient. In Fig.~\ref{fig-inner} we 
show the effect of inner loop
residue on ``pseudoscalar'' propagators for heavy
quarks. The curves are slightly shifted for
clarity. For $ma=0.450$, even an inner loop residue of $10^{-6}$
appears to be good enough. However, for a larger $ma=0.630$, this
residue is not good enough for $t>$30. For our production runs, we
choose an inner loop residue of $10^{-8}$ and outer loop residue of
$10^{-5}$. We have tested outer loop residue of $10^{-7}$, two orders
of magnitude better. This 
affects results at less than half percent level, so we deem outer loop
residue of
 $10^{-5}$ to be sufficient. This residue of $10^{-5}$ is demanded for
the lightest quark mass. Near the charm mass, the residue obtained through the multi-mass
inversion algorithm is $\approx 2
\times 10^{-9}$.

\vspace{-0.5cm}
\section{Analysis}

In this paper, we study five charmonium states shown in
Table~\ref{table-charm-states} -- $\eta_c(^1S_0)$, $J/\Psi(^3S_1)$,
$h_c(^1P_1)$, $\chi^0_c(^3P_0)$ and
$\chi_c^1(^3P_1)$. 
For the $P$ states, there are two possible operators --
one (denoted by $\Gamma$) simply using appropriate $\gamma$ matrices and the
other (denoted by $\Delta$) using a
derivative as well as $\gamma$ matrices. We always use a $\Gamma$ operator for the source,
because using a 
$\Delta$ operator for source would require additional inversions. (It is
for this reason we do not study $\chi_c^2$. This state has no $\Gamma$
operator.) 
Using a $\Delta$ sink does not cost additional
inversions. Thus for our P state analysis, we have three
possibilities -- $\Gamma$, $\Delta$ or $\Gamma\Delta$. The last one is our
notation for a simultaneous fit to both $\Gamma$ and $\Delta$ sink
correlators.

\begin{table}[h]
\begin{center}
\begin{tabular}{|c|c|c|c|c|c|}
\hline
\vspace{-0.3cm}
&&&&&\\
& $^{2s+1}L_J$ & $J^{PC}$& field $\Gamma$ & field $\Delta$ & mass  \\
&&&&&(GeV)\\
\vspace{-0.3cm}
&&&&&\\
\hline
\vspace{-0.3cm}
&&&&&\\
$\eta_c$    & $^1S_0$ & $0^{-+}$ & $\bar{\psi} \gamma_5 \psi$ & --- &
2.979 \\
$J/\psi$    & $^3S_1$ & $1^{--}$ & $\bar{\psi} \gamma_{\mu} \psi$ & ---
& 3.097 \\
$h_c$       & $^1P_1$ & $1^{+-}$ & $\bar{\psi} \sigma_{ij} \psi$ &
$\bar{\psi} \gamma_5 \Delta_i \psi$ & 3.526 \\
$\chi_{c0}$ & $^3P_0$ & $0^{++}$ & $\bar{\psi} \psi$ & $\bar{\psi}
\Sigma_i \gamma_i \Delta_i \psi$ & 3.417 \\
$\chi_{c1}$ & $^3P_1$ & $1^{++}$ & $\bar{\psi} \gamma_i \gamma_5 \psi$ &
$\bar{\psi} \left( \gamma_i \Delta_j - \gamma_j \Delta_i \right) \psi$ &
3.511 \\
\hline
\end{tabular}
\vspace{0.3cm}
\caption{Charmonium states. For the $P$ states, there are two possible
interpolating fields, denoted by $\Gamma$ and $\Delta$. Experimental masses in GeV
are shown.  \label{table-charm-states}}
\end{center}
\end{table}

\begin{figure}[h]
\begin{center}
\includegraphics[width=12cm]{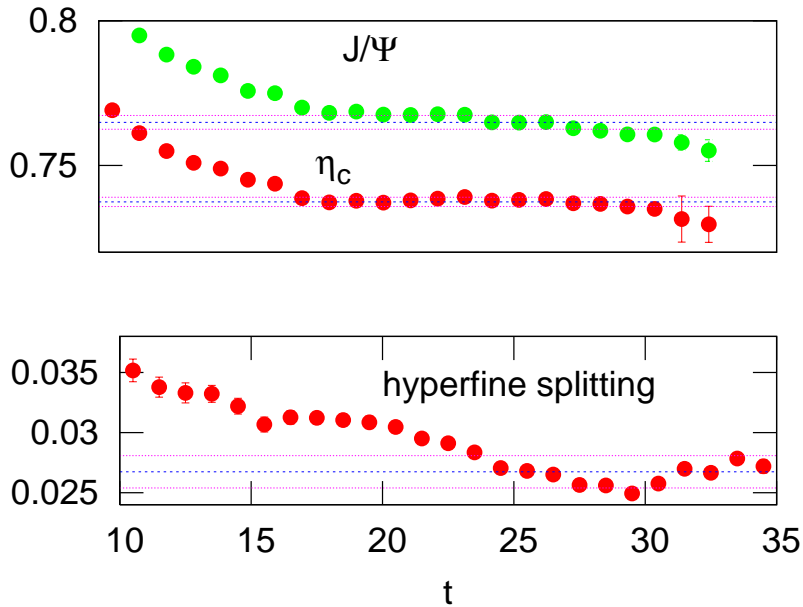}
\end{center}
\caption{Effective masses for the pseudoscalar and vector correlators.
The plateau for the ratio of vector to pseudoscalar correlator is also
shown. We use this ratio to obtain our results for hyperfine splitting.
These plots are for $ma$ = 0.35. \label{fig-eff-S}}
\end{figure}

\begin{figure}[h]
\begin{center}
\includegraphics[width=12cm]{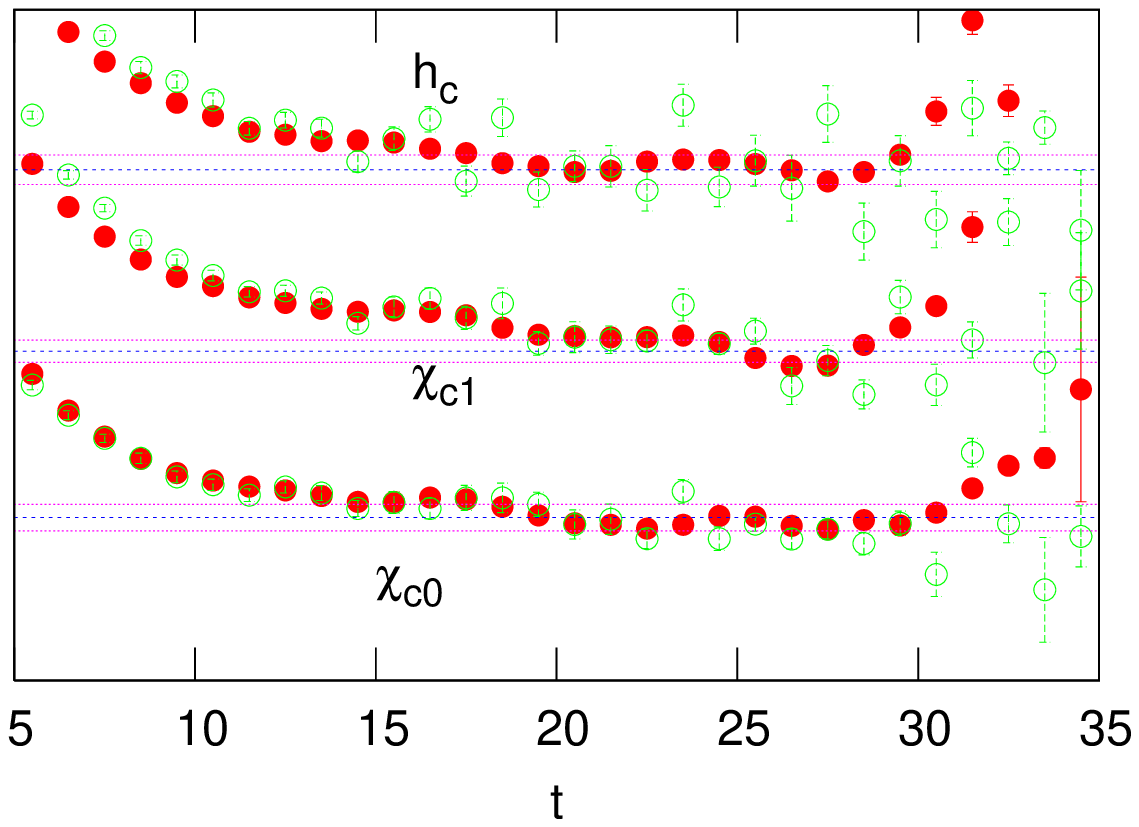}
\end{center}
\caption{Effective masses for the P states. The filled circles correspond
to $\Gamma$ operator and the open circles to $\Delta$ operator. The
plots for different P states are shifted along the y-axis. These effective
masses are rather noisy, and we use conservative error bars on our
fitted results. Again, the  plots are for $ma$ = 0.35.  \label{fig-eff-P}}
\end{figure}

The effective mass
plots for the pseudoscalar and the vector states are shown in
Fig.~\ref{fig-eff-S}. The lower half of this figure shows the effective
hyperfine splitting from the ratio of vector to pseudoscalar correlators.
These show a long plateau to justify a single exponential
fit. 
For the $P$ states,
the effective masses are shown in Fig.~\ref{fig-eff-P}. These have much larger error bars, but they are still
flat. The data gets noisy beyond $t=30$ and precision problems cannot
be excluded for channels other than the pseudoscalar meson. We do not
use time-slices beyond 30 in our fits.

We use two ways to set the scale -- from 
$r_0$ (using 0.5~fm) and from the ($1\bar{P}-1\bar{S}$)
splitting in the charmonium system. 
The singlet $P$ mass $m_{h_c}$ is used for $\bar{P}$,
and $(3 m_{J/\psi} + m_{\eta_c})/4$ for $\bar{S}$ mass.
The ($1\bar{P}-1\bar{S}$) scale analysis has three sub-cases,
depending on which of $\Gamma$, $\Delta$ or $\Gamma\Delta$ fit is used
for $h_c$.

\begin{figure}[h] 
\begin{center}
\includegraphics[width=12cm]{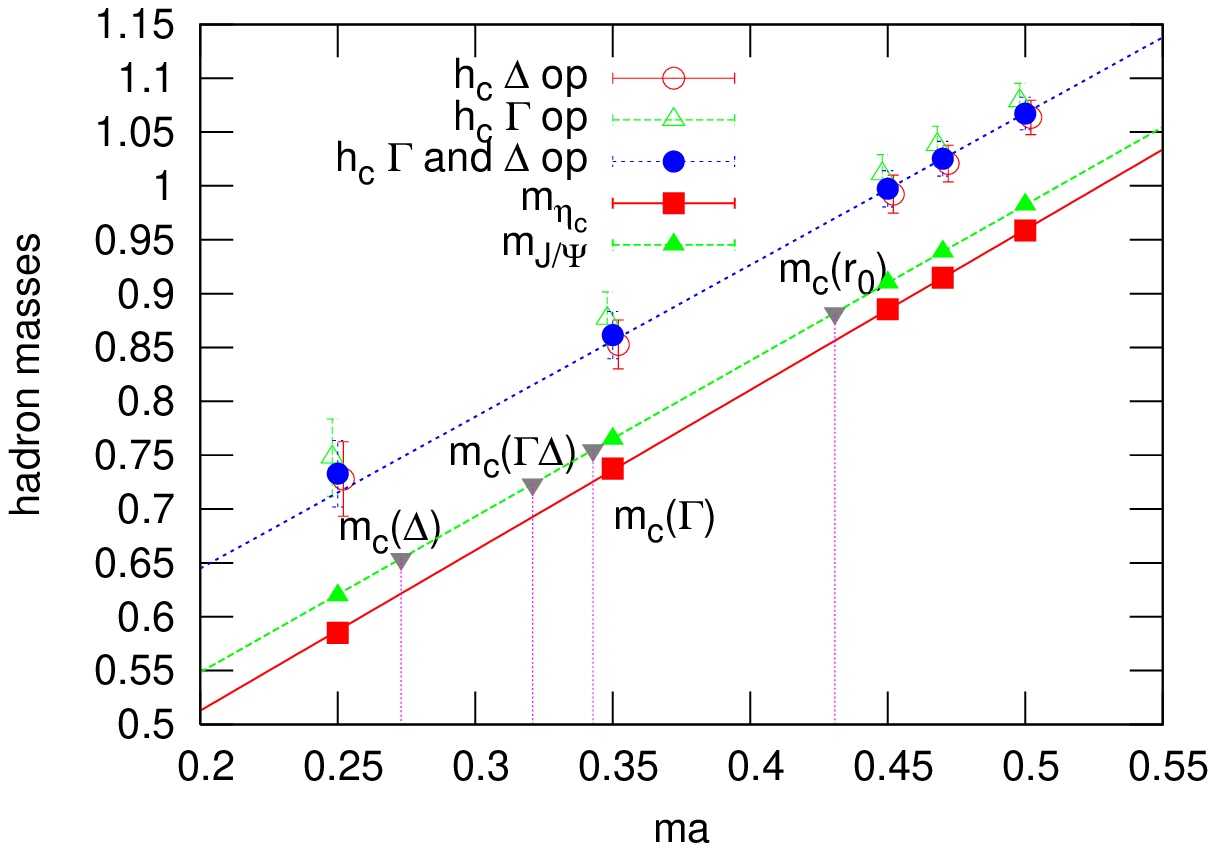} 
\end{center}
\caption{We fit the meson masses linearly in quark mass.  Fits are shown for 
$\eta_c,J/\psi$ and $h_c$ masses. All masses are in lattice units. $h_c$
masses obtained using $\Gamma$ and $\Delta$ operators are also shown,
but the fit line is shown only for the $\Gamma\Delta$ fit. \label{fig-inter}} 
\end{figure}

\begin{figure}[h]
\begin{center}
\includegraphics[width=12cm]{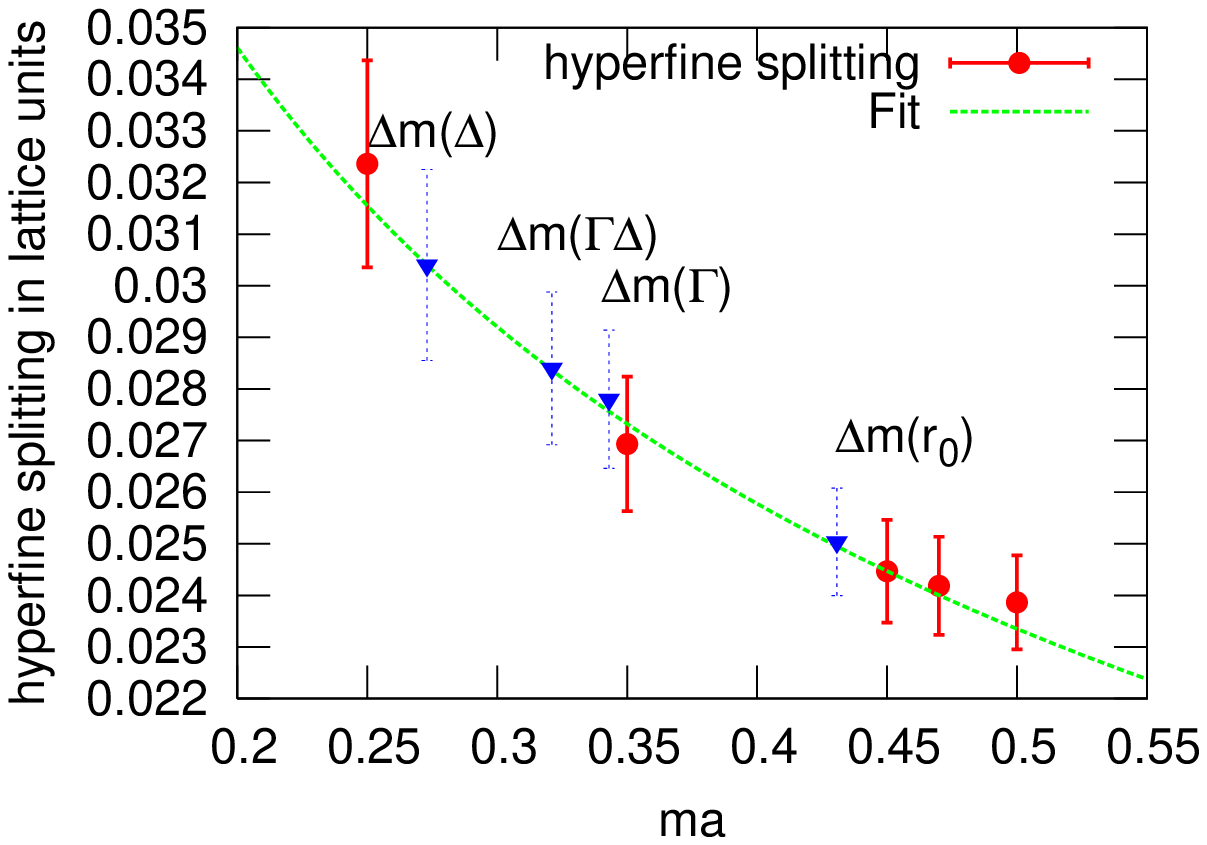}
\end{center}
\caption{Hyperfine splitting as a function of quark mass, with
interpolation shown at $m_ca$.\label{fig-hyp}}
\end{figure}

\begin{figure}[h]
\begin{center}
\includegraphics[width=12cm]{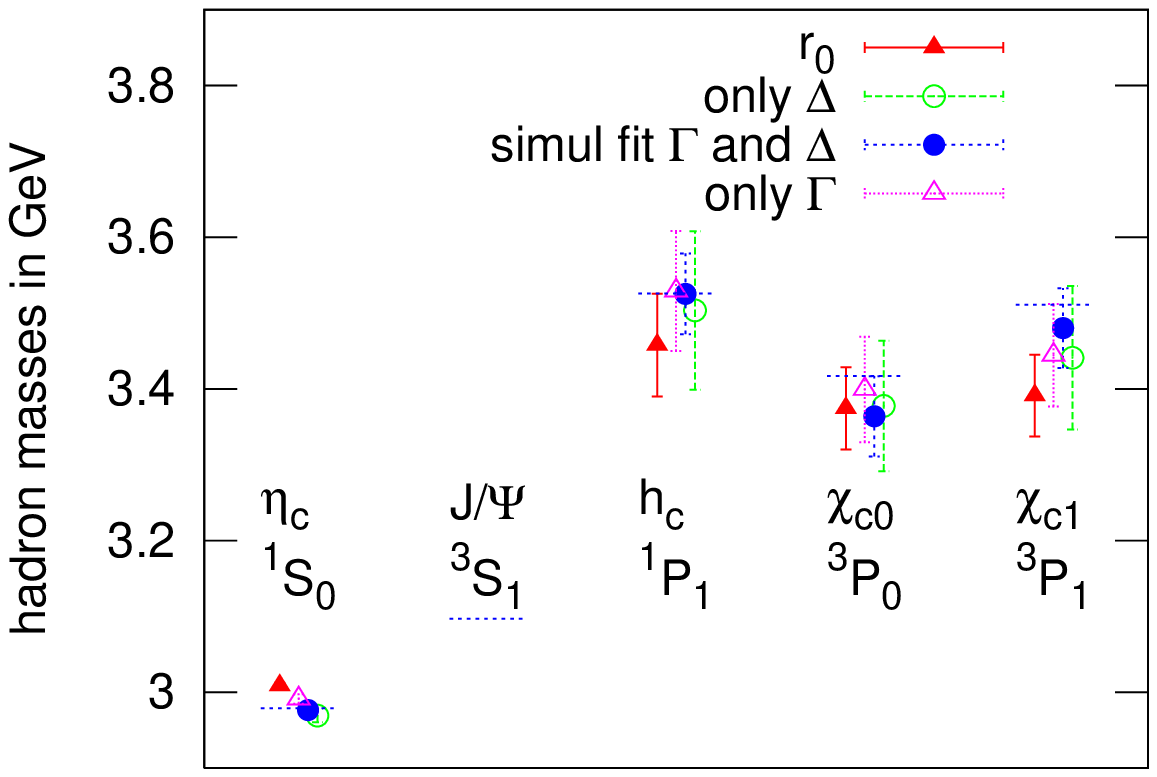}
\end{center}
\caption{Charmonium spectrum in physical units. Results from both $r_0$
and
$1\bar{P}-1\bar{S}$ scales are shown. Note, for the latter scale, a
linear combination of
$h_c$ and $\eta_c$ masses, along with the $J/\psi$ mass, is used for
input.\label{fig-spectrum}}
\end{figure}

We present the $r_0$ results first. 
The lattice spacing for the $\beta$ we use is 0.0561 fm
\cite{sommer}. The experimental $m_{J/\psi}$ is used to set $m_c$ (in
lattice
units). Interpolation for $m_{J/\psi}$ as a function of $ma$ is shown in
Fig.~\ref{fig-inter}. A straight line fit is used. Interpolation for the hyperfine splitting is shown in
Fig.~\ref{fig-hyp}.
The fit
form used is $(m_{J/\psi}-m_{\eta_c})a = A/\sqrt{ma}+B/ma$~\cite{liu-wong}. Knowing the
charm mass and the scale, the hyperfine splitting in
MeV can be determined. Our result for the hyperfine splitting using
$r_0$ scale is 88(4) MeV. This is considerably higher than the quenched
results from Wilson-type fermions. The spectrum obtained using $r_0$
scale is shown in Fig~\ref{fig-spectrum}. The corresponding results can
be found in Table~\ref{table-spectrum}.

The ($1\bar{P}-1\bar{S}$) scale has the advantage that it is set within the
charmonium system, using masses of physical particles, so it is expected 
to be more relevant for this system, and it is model independent. 
However, we have large errors on the $P$ states. Consequently, the scale set from
($1\bar{P}-1\bar{S}$) splitting itself will have about 12\% error, which
is not included in
the direct statistical errors on various masses quoted below. The interpolation for
the $\Gamma\Delta$ fit for $m_{h_c}$ is shown in Fig.~\ref{fig-inter},
along with the interpolations for $m_{J/\psi}$ and $m_{\eta_c}$. We also
show $m_{h_c}$ obtained using $\Gamma$ and $\Delta$ fits on the same
plot. It is clear from this plot that $m_{h_c}$ obtained from the three
fits completely agree within error bars. However, the slight difference
in $m_{h_c}$ in the three cases changes the scale, the charm mass and
the hyperfine splitting values considerably. 

In the case of the spin splitting scale, the determination of $a$
and
$m_ca$ is entangled. The procedure we follow to disentangle these is as
follows.
As shown in Fig.~\ref{fig-inter} all hadron masses in lattice units are
fitted to a straight line,
$m_ha = A_h.ma + B_h$.
Lattice spacing $a$ and bare charm quark mass
$m_ca$ are
two
unknowns; $m_{J/\psi}$ and
$m(1\bar{P}-1\bar{S})$ in physical units are the
two inputs. We solve for $a$ and $m_ca$ to obtain values shown in Table
\ref{table-spectrum}.
The charm masses obtained
are indicated in Fig.~\ref{fig-inter}. We would like to point out that while $m_ca$
in lattice units differs considerably in the three sub-cases of
($1\bar{P}-1\bar{S}$) analysis, values for $m_c$ in GeV, tabulated in Table
\ref{table-spectrum}, cluster much tighter.

The value we obtain for the hyperfine splitting in MeV is extremely
sensitive to the value used for the lattice spacing $a$. For a slightly
smaller $a$, the hyperfine splitting in lattice units is considerably
larger, since it falls rapidly with increasing $a$, as seen in Fig.~\ref{fig-hyp}. Converting this to
physical units further increases the value. As a result, our results
from the three sub-cases of ($1\bar{P}-1\bar{S}$) analysis look quite
different -- 113(5) MeV using $\Gamma$, 121(6) MeV using $\Gamma\Delta$
and 144(9) MeV using $\Delta$. We would like to emphasize here that the
errors quoted are only direct statistical errors, and the errors on $a$
are large enough to bring these results in statistical agreement with each other.

Fig.~\ref{fig-spectrum} shows the charmonium spectrum obtained from both
$r_0$ and $1\bar{P}-1\bar{S}$ analysis. Agreement with the experimental
values is much better for the $1\bar{P}-1\bar{S}$ scale. The 
agreement with experimental numbers for all the particles studied is
very reasonable, indicating that the discretization
errors must indeed be small for overlap fermions. This is because the
different mass differences are supposed to measure differently defined
quark masses $M_2$, $M_E$, etc. \cite{fermilab}. The inequality of these quark masses
implies discretization errors. If all the mass differences come out
right, it would imply that $M_1 \approx M_2 \approx M_E$, and small discretization
errors.

Finally we summarize the results in Table \ref{table-spectrum}.
The errors quoted are only statistical; the error on $a$ is not included. 
All masses are in
GeV. Our value for the hyperfine splitting using $1\bar{P}-1\bar{S}$
scale, and 
simultaneous fits to $\Gamma$ and $\Delta$ correlators 
actually agrees with experiment. This is fortuitous, because
the contribution from dynamical fermions is not included, and may be
significant. However, there is no real contradiction here, because we
have substantial statistical and systematic errors, as detailed below:

\begin{table}[h]
\begin{center}
\begin{tabular}{|c|c|c|c|c|c|}
\hline
\vspace{-0.3cm}
&& \multicolumn{3}{c|}{ }&\\
&$a(r_0)$ & \multicolumn{3}{c|}{$a(1\bar{P}-1\bar{S})$} & Expt \\ \cline{3-5}
\vspace{-0.3cm}
&&&&&\\
&&$\Gamma$&$\Gamma\Delta$&$\Delta$& \\
\hline
\vspace{-0.3cm}
&&&&&\\
$\eta_c$&3.017(4)&2.977(6)&2.967(7)&2.943(9)&2.980\\
$J/\psi$&---&---&---&---&3.097\\
$J/\psi-\eta_c$&0.088(4)&0.113(5)&0.121(6)&0.144(9)&0.117\\
$h_c$&3.44(7)\ \ &3.53(8)\ \ &3.49(9)\ \ &3.47(12)\ &3.526\ \\
$\chi_{c0}$&3.36(5)\ \ &3.41(7)\ \ &3.43(8)\ \ &3.39(10)\ &3.41\ \ \\
$\chi_{c1}$&3.39(5)\ \ &3.46(7)\ \ &3.41(7)\ \ &3.45(10)\ &3.511\ \\
\hline
\vspace{-0.3cm}
&&&&&\\
$m_ca$&0.431&0.343&0.321&0.273&---\\
$m_c$(GeV)&1.52&1.41&1.38&1.30&---\\
$a(fm)$&0.0561&0.0480&0.0460&0.0416&---\\
\hline
\end{tabular}
\vspace{0.3cm}
\caption{Charmonium spectrum (GeV). Only direct statistical errors are
included; the statistical error on the lattice spacing $a$ and
systematic errors are not included in this table. \label{table-spectrum}}
\end{center}
\end{table}

\begin{enumerate}
\item{Direct statistical errors: These are quoted in 
Table~\ref{table-spectrum}.}
\item{Statistical error on $a$: In the ($1\bar{P}-1\bar{S}$) scale, this is
primarily due to the error on
$h_c$ mass, which is about 53 MeV. This is about 12\% of the physical
($1\bar{P}-1\bar{S}$) mass difference of 458 MeV. Note, this error is
absent when the scale is set using $r_0$. On the other hand, $r_0$ is a
model dependent scale, and it can have comparable errors. It has been
pointed out that 0.45~fm may be a better value to use for $r_0$ than
0.50~fm \cite{lepage}. Using this value brings our $r_0$ results closer
to the ($1\bar{P}-1\bar{S}$) results.}

\item{Discretization errors: As explained in Section 1, this is
estimated at about 5\%, from the dispersion relation.}
\item{Finite volume errors: Our simulations are performed on a box size
of only 0.8~fm, hence it is not inconceivable that the $P$ states have
some finite volume errors. However, even this small box should be large
enough for the $S$ state particles $---$ $J/\Psi$ and
$\eta_c$.}
\item{Quenched approximation: Dynamical fermions are expected to
increase the value of hyperfine splitting. This increase is about 20 MeV
for the Wilson-type fermions~\cite{dyn}. On the other hand, a study with
NRQCD~\cite{NR-dyn} does not find significant contribution due to dynamical
fermions.}
\item{Exclusion of OZI-suppressed diagrams: While a contribution of
about 20 MeV cannot be ruled out, the contribution due to these appears
to be small in the charm quark region~\cite{disconn}. 
Lattice calculations with smaller
statistical and systematic errors are needed to settle this issue.}
\end{enumerate}

\section{Summary}
We have presented the first study of the charmonium spectrum using
overlap
fermions. We get a better agreement with the experimental spectrum using
$1\bar{P}-1\bar{S}$ scale rather than the $r_0$ scale.
Our value for hyperfine splitting is 121(6)~MeV and 88(4)~MeV
using $1\bar{P}-1\bar{S}$ and $r_0$ scale respectively.
This is considerably higher than the quenched clover
results. This conclusion cannot be escaped even if it is argued that
our $P$ state results are affected by finite volume errors. Unquenched
overlap results with more statistics and somewhat larger box size may
very well settle the charmonium hyperfine splitting issue.
                                                                                
This work is supported in part by U.S. Department of Energy
under grants DE-FG05-84ER40154 and DE-FG02-95ER40907. The computing
resources at
NERSC (operated by DOE under DE-AC03-76SF00098) are also acknowledged.
Y. Chen and S. J. Dong are partly supported by NSFC  \\
(\#10235040 and
\#10075051 )

\clearpage


\end{document}